\documentclass{article}

\PassOptionsToPackage{numbers}{natbib}
\usepackage[preprint]{neurips_2024}

\usepackage[utf8]{inputenc} 
\usepackage[T1]{fontenc}    
\usepackage{hyperref}       
\usepackage{url}            
\usepackage{booktabs}       
\usepackage{amsfonts}       
\usepackage{nicefrac}       
\usepackage{microtype}      
\usepackage{xcolor}         

\usepackage{wrapfig}
\usepackage{picinpar}
\usepackage{subfigure}
\usepackage{caption}
\usepackage{subcaption}
\usepackage{amsthm}
\usepackage{amsmath}
\usepackage{mathtools}
\usepackage{amsfonts}
\usepackage{amssymb}
\usepackage{bm}
\usepackage{algorithm}
\usepackage{algpseudocode}
\usepackage[pdftex]{graphicx} 
\usepackage[normalem]{ulem}

\title{
Mind's Eye: Image Recognition by EEG via Multimodal Similarity-Keeping Contrastive Learning
}

\author{%
  Chi-Sheng Chen \\
  Department of Computer Science\\
  National Yang Ming Chiao Tung University \\  
  Hsinchu, Taiwan\\
  \texttt{m50816m50816@gmail.com} \\
  \And
  Chun-Shu Wei \\
  Department of Computer Science\\
  National Yang Ming Chiao Tung University\\
  Hsinchu, Taiwan \\
  \texttt{wei@nycu.edu.tw} \\
}

\begin{document}
\maketitle

\begin{abstract}
Decoding images from non-invasive electroencephalographic (EEG) signals has been a grand challenge in understanding how the human brain process visual information in real-world scenarios. To cope with the issues of signal-to-noise ratio and nonstationarity, this paper introduces a MUltimodal Similarity-keeping contrastivE learning (MUSE) framework for zero-shot EEG-based image classification. We develop a series of multivariate time-series encoders tailored for EEG signals and assess the efficacy of regularized contrastive EEG-Image pretraining using an extensive visual EEG dataset. Our method achieves state-of-the-art performance, with a top-1 accuracy of $19.3\%$ and a top-5 accuracy of $48.8\%$ in 200-way zero-shot image classification. Furthermore, we visualize neural patterns via model interpretation, shedding light on the visual processing dynamics in the human brain. The code repository for this work is available at: \url{https://github.com/ChiShengChen/MUSE_EEG}.
\end{abstract}

\section{Introduction}

Understanding visual processing in the human brain remains a profound challenge at the intersection of neuroscience and artificial intelligence. Visual processing involves a complex sequence of neural mechanisms across various brain regions, enabling the intricate processing of visual stimuli \cite{riesenhuber1999hierarchical, miyawaki2008visual, liu2009timing, dicarlo2012does, gifford2022large}. The development of deep learning techniques, such as convolutional neural networks (CNNs), has been significantly inspired by our understanding of these neural mechanisms \cite{fukushima1980neocognitron, lecun1998gradient, lecun2015deep}. Unveiling the brain dynamics of visual processing in real-world contexts holds the potential to inspire future advancements in artificial intelligence (AI), continuing the cycle of innovation driven by biological insights \cite{hassabis2017neuroscience, ullman2019using}. Recent studies have advanced our understanding of visual processing in the human brain through the observation of brain activity using various neuromonitoring modalities \cite{he2011electrophysiological}. Electroencephalography (EEG), as a non-invasive, portable modality with high-temporal resolution, offers a unique window into visual processing by revealing the instantaneous neural dynamics of visual perception and recognition in real-world contexts \cite{rousselet2007single, samaha2015speed, wei2023towards}.

Decoding images from EEG signals represents a promising approach to study the mechanisms of visual processing. By leveraging EEG, researchers can gain insight into the temporal evolution of neural responses to visual stimuli \cite{robinson2017very}. However, this endeavor faces significant obstacles, primarily due to the low signal-to-noise ratio and nonstationarity of EEG signals \cite{kaplan2005nonstationary, uriguen2015eeg}. Addressing these challenges is crucial for advancing our understanding of visual cognition and for developing robust EEG-based image decoding or brain-computer interfacing (BCI) systems.
Early studies in EEG-based image decoding have been constrained by the use of small datasets, limiting their ability to develop generalizable models \cite{spampinato2017deep, tirupattur2018thoughtviz}. More recent work has utilized larger datasets collected through the rapid serial visual presentation (RSVP) paradigm, where images are presented in quick succession to elicit brain responses \cite{gifford2022large, song2023decoding}. Despite these advances, the performance of existing methods remains suboptimal, underscoring the need for dedicated design of EEG encoding network architectures that consider the brain’s mechanisms and EEG characteristics.

To address the challenges in EEG-based image decoding, we present a novel self-supervised framework, coined as multimodal similarity-keeping contrastive learning (MUSE), dedicated to cross-modality contrastive learning between EEG and image data. We develop a series of multivariate time-series encoder network architectures tailored for EEG processing that facilitate the cross-modality contrastive learning with an advanced off-the-shelf image encoder (CLIP-ViT \cite{radford2021learning}). These encoders feature an upstream spatial convolution of EEG data for the sake of feature extraction and noise suppression \cite{wei2019spatial, pan2022matt}. Additionally, we propose an innovative similarity-keeping contrastive learning mechanism, inspired by the cortical mapping organization of visual object representation in the inferotemporal (IT) cortex \cite{bao2020map}, to regularize the contrastive learning process using the information of inter-object relationships within both EEG and image samples.

Furthermore, we employ model interpretation techniques to visualize the neural patterns of image processing, offering a deeper understanding of the underlying dynamics of visual cognition in the human brain. The contributions of this work are threefold:

\begin{itemize}
\item We introduce a novel self-supervised multimodal similarity-keeping contrastive learning (MUSE) framework that achieves state-of-the-art performance in zero-shot EEG-based image recognition.
\item We propose EEG encoders with upstream spatial convolution and similarity-keeping regularization to enhance EEG-image cross-modality contrastive learning.
\item We visualize neural patterns through model interpretation to provide neuroscientific insights into the spatial and temporal brain dynamics of visual processing.
\end{itemize}

\begin{figure}
    \centering
    \includegraphics[width=1\linewidth]{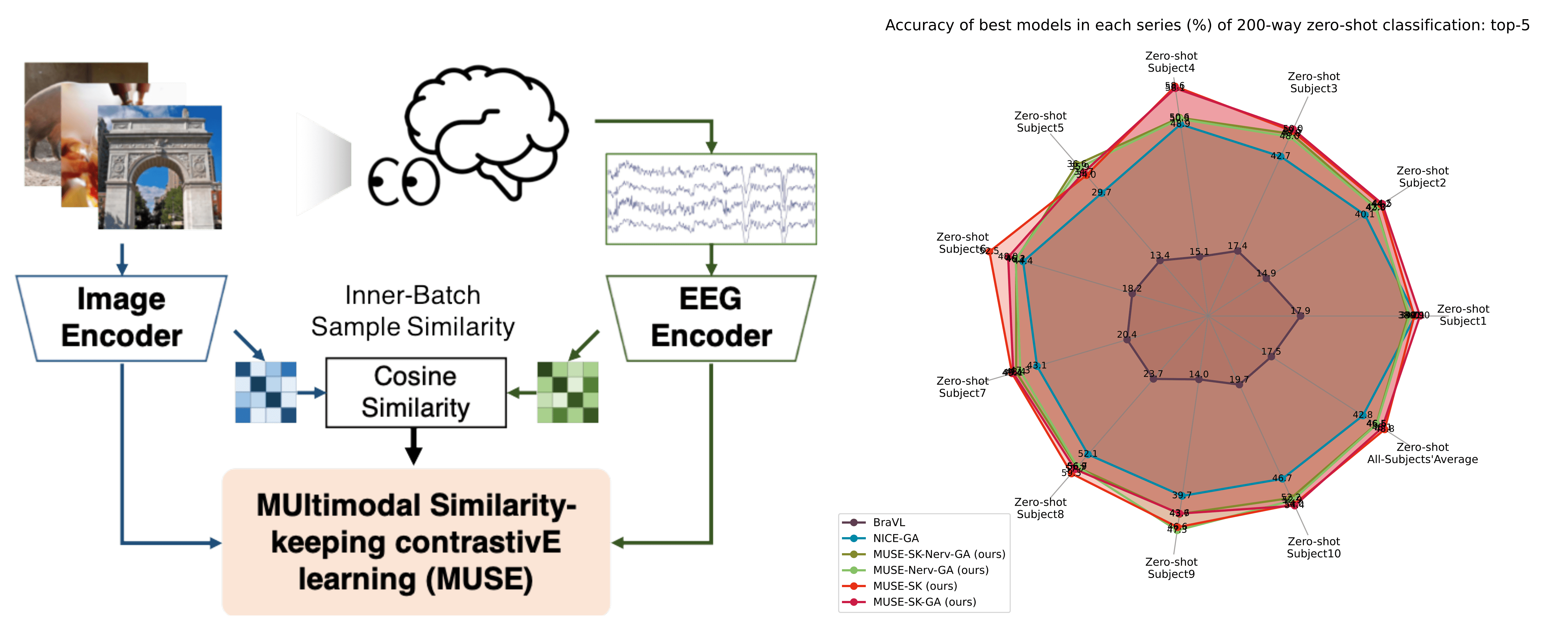}
    \caption{Schematic illustration of the proposed MUltimodal Similarity-keeping contrastivE learning (MUSE) framework. During the training phase, EEG-image pairs are independently processed by an EEG encoder and an image encoder. The objectives of the MUSE framework are twofold: 1) maximize the separation between matched and unmatched pairs, and 2) maintain the inner-batch sample similarity within each EEG-image pair (see Algorithm \ref{alg:1} for details). In the test phase, an unseen EEG sample is passed through the EEG encoder, which identifies the most similar image from a set of unseen images based on cross-modality embedding similarity.}
    \label{fig:ga}
\end{figure}

\begin{figure}
    \centering
    \includegraphics[width=.9\linewidth]{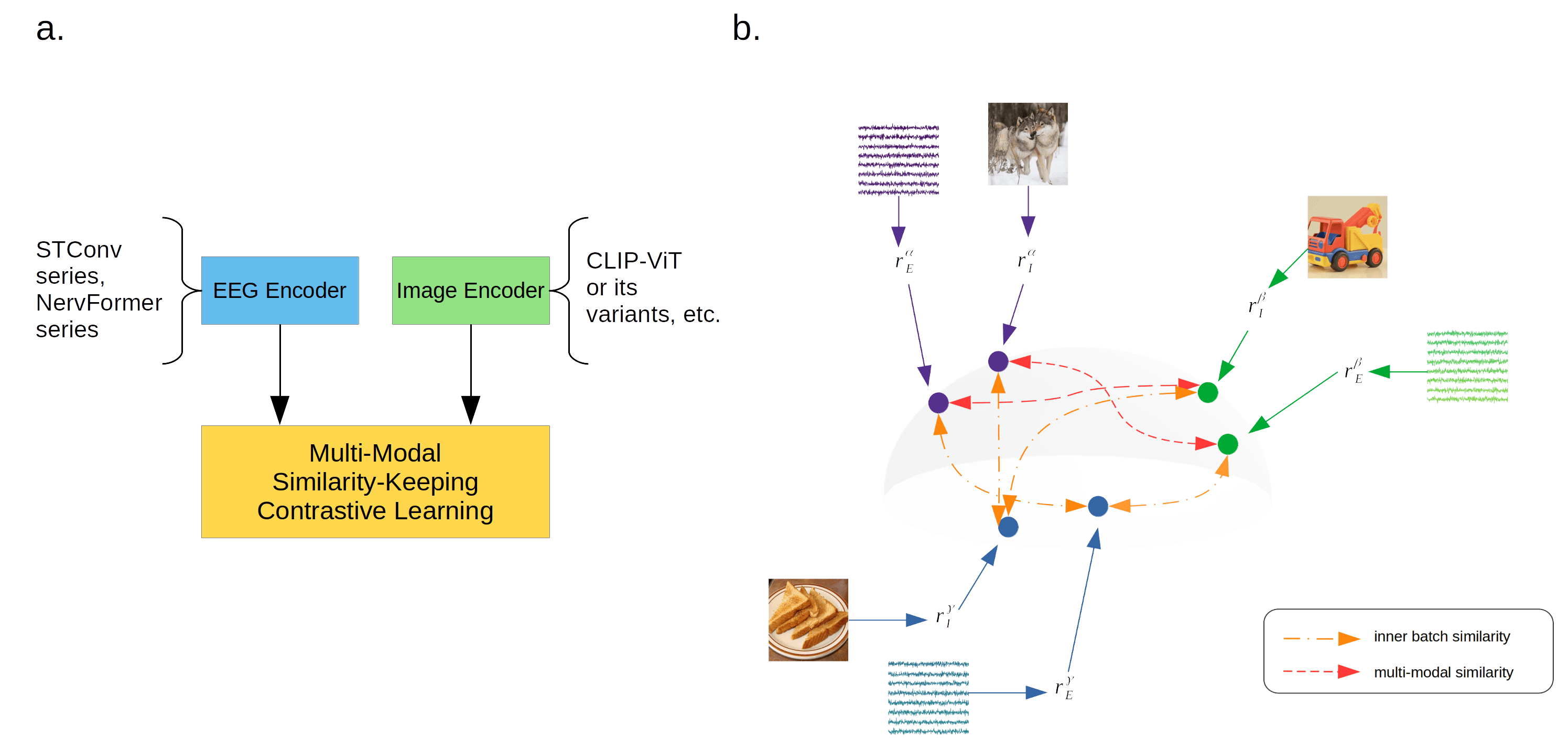}
    \caption{(a.) The whole view of this work. (b.) Illustration on feature space of multimodal similarity-keeping contrastive learning framework (MUSE), different from traditional contrastive learning only focus on multimodal similarity, MUSE both consider the multimodal similarity and inner batch similarity in the loss function. \(r\) denotes representation. \(I\) and \(E\) denotes image and EEG signal, respectively.}
    \label{fig:music-plot}
\end{figure}



\section{Related Works}
\label{gen_inst}

\subsection{Decoding Visual Information from Brain Signals}

Interpreting visual data from the human brain has been a longstanding challenge at the intersection of neuroscience and computer science \cite{riesenhuber1999hierarchical, miyawaki2008visual, dicarlo2012does, gifford2022large}. Despite significant advancements in understanding static visual inputs, rapidly and accurately extracting meaningful information from natural imagery remains difficult \cite{kay2008identifying, chen2023ssvep}. Previous efforts have primarily utilized functional magnetic resonance imaging (fMRI) \cite{mai2023brain, takagi2023high, scotti2024mindeye2}, which has demonstrated the ability to capture meaningful content and structural details from visual processing in the brain. However, fMRI relies on detecting changes in blood oxygenation, resulting in a temporal lag of several seconds per stimulus, thereby limiting its utility for real-time applications. Additionally, fMRI is expensive and requires large, stationary equipment.

In contrast, electroencephalography (EEG) offers superior temporal resolution, immediate data feedback, and portable, cost-effective hardware. These attributes position EEG as a promising candidate for personal brain-computer interface technology. Nevertheless, current methods for using EEG to extract semantic information for image classification have not achieved satisfactory results \cite{ahmed2021object, 10073607, song2023decoding}, highlighting the need for improved approaches. Previous methodologies have often relied on supervised learning techniques with a limited set of image categories, ignoring the intrinsic correlations between visual stimuli and neural responses \cite{10073607, spampinato2017deep, singh2024learning}. These limitations impair their effectiveness in real-world scenarios that require the generalization to recognize novel, unfamiliar object categories. To address these issues, \cite{du2023decoding} first attempted zero-shot classification using the largest available EEG-image database \cite{gifford2022large} with a multilayer MLP and joint EEG-image-text representation, while \cite{song2023decoding} employed a contrastive learning method. However, \cite{song2023decoding} utilized a basic contrastive learning framework based on CLIP \cite{radford2021learning}. Our work improves upon this framework and the EEG encoder, introducing a self-supervised learning approach for EEG-based image decoding. This framework allows the model to generalize to object recognition tasks without specific prior training, demonstrating its effectiveness.

\subsection{Multimodal Contrastive Learning}

In recent years, after the success of the traditional contrastive learning models on the same modal data like text and image \cite{tian2020contrastive, he2020momentum, grill2020bootstrap, chen2020simple}, the development of multimodal contrastive learning has reached significant advancements in the field of self-supervised learning, particularly in tasks that contain the integration of multiple types of data. This method leverages the strengths of various modalities (e.g., text, images, video) to boost model generalization across diverse datasets. Multimodal contrastive learning aligns representations from different modalities within a shared embedding space, facilitating robust, modality-invariant feature learning. This enhances capabilities in cross-modal retrieval and zero-shot learning. Typically, a two-tower network architecture processes each modality independently, with outputs converging in the embedding space where contrastive loss minimizes distances between similar pairs and maximizes distances between dissimilar ones.
One of the most popular and successful multimodal contrastive learning framework is CLIP \cite{radford2021learning}, which project both the image and text in to the same feature space. Nevertheless, because datasets containing both time-series signals like EEG and image data are quite rare, there has been little research applying contrastive learning methods to this combination of temporal and visual information. To our best knowledge, \cite{ye2022see} is maybe the first work introduced the EEG-image contrastive learning on obtaining the EEG-image representation for image reconstruction downstream task but do not do the zero-shot classification. \cite{singh2024learning} introduced the EEGClip network for joint representation learning between EEG signal and image but it just do supervised learning. \cite{song2023decoding} first try to design the EEG encoder on EEG-image contrastive learning , but the work only modified the encoders. This area remains largely uncharted and calls for new, specialized contrastive learning techniques to handle these joint time-series and image modalities effectively.


\begin{figure}
    \centering
    \includegraphics[width=.9\linewidth]{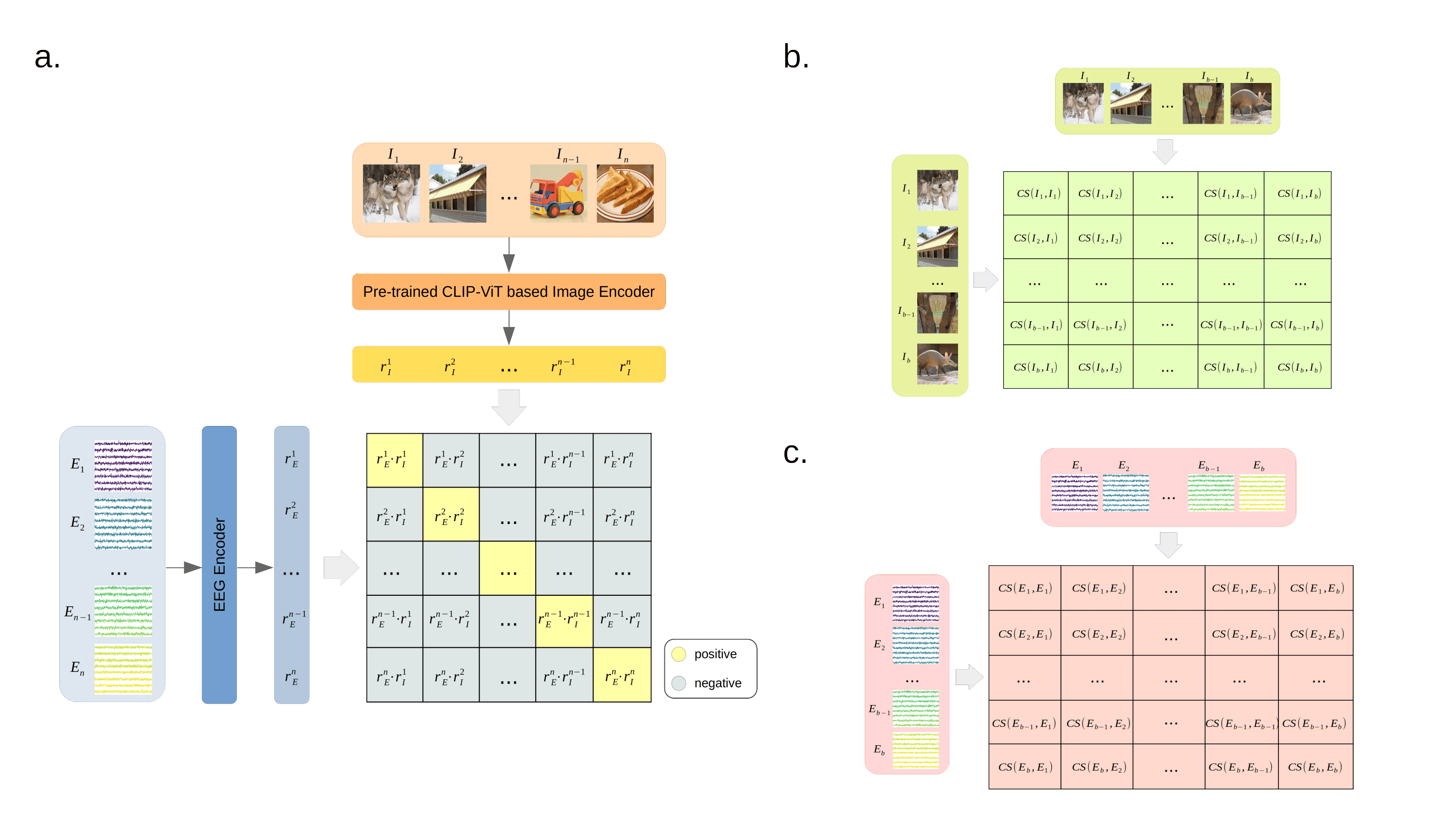}
    \caption{The details of the MUSE. (a.) The contrastive learning loss is calculated from EEG encoding and image encoding. (b.)(c.) The similarity-keeping loss comes from the final similarity of self-batch similarity of the input modal data.}
    \label{fig:sim_plot}
\end{figure}

\begin{figure}
    \centering
    \includegraphics[width=.9\linewidth]{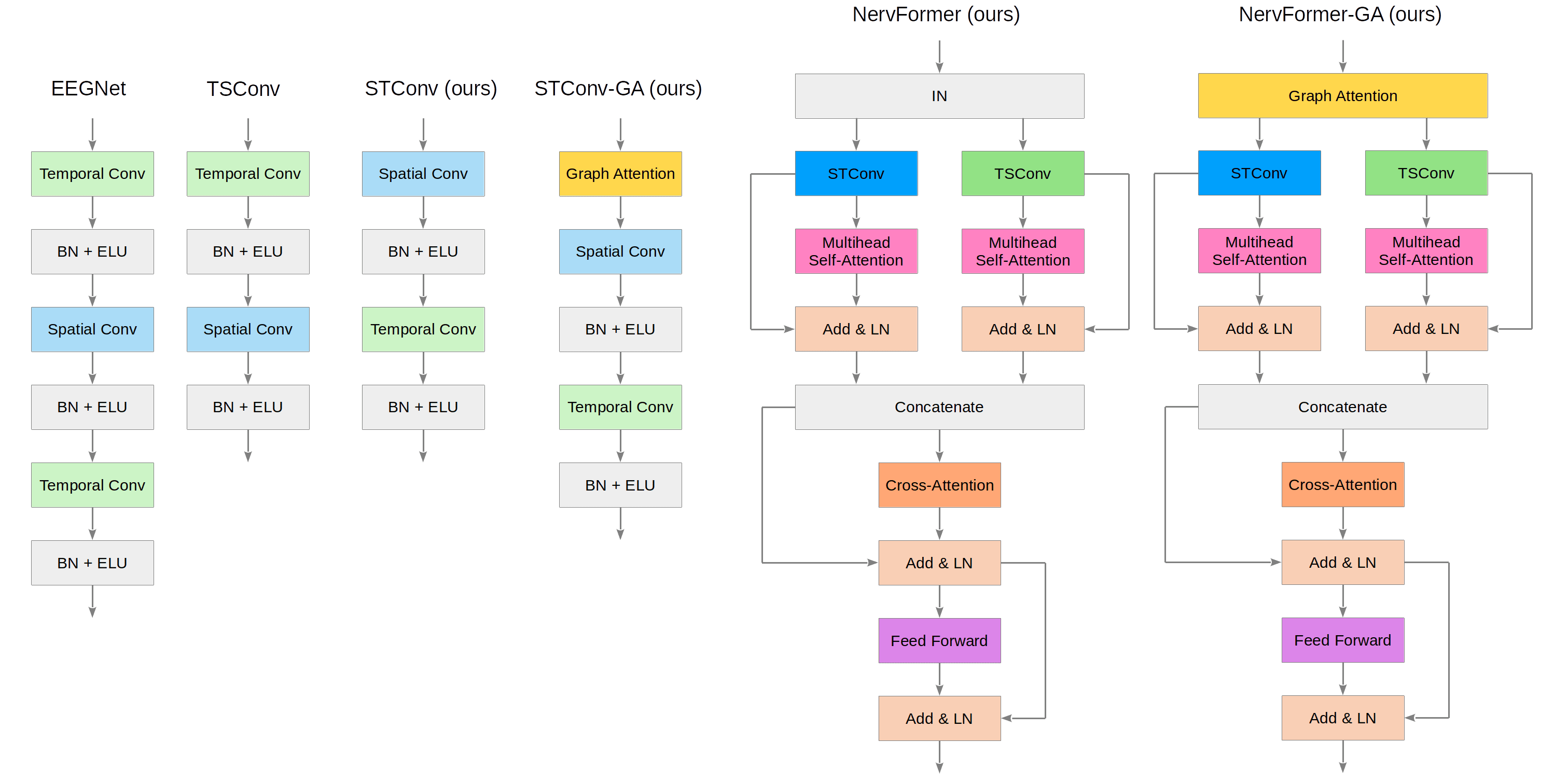}
    \caption{The model structure comparison. Where BN denotes batch normalization, IN denotes instance normalization, LN denotes layer normalization, respectively.}
    \label{fig:model_compare}
\end{figure}
\section{Methodology}
\subsection{Overview}
This section introduces the Multimodal Similarity-Keeping Contrastive Learning (MUSE) framework, comprising the EEG encoder, image encoder, and the contrastive learning method. Our contribution encompasses cutting-edge EEG encoders tailored for zero-shot classification tasks: the Spatial-Temporal convolution (STConv) and NervFormer architectures, along with a pioneering regularized contrastive learning approach featuring a novel similarity-keeping loss.

\begin{algorithm}[t!]
\caption{Multimodal Similarity-Keeping Contrastive Learning framework (MUSE)} 
\label{alg:1}
\small
{\fontfamily{cmr}\selectfont
\begin{algorithmic}[1]
    \State \textbf{Input}: (Image, EEG)  
    \Comment{stimulus \& response}
    \State \textbf{Model}: $Enc_{img}$: CLIP-ViT or its variance, $Enc_{eeg}$: STConv or NervFormer
    \Statex 
    \State \# E : \((batch, channel, electrode, data \ sample)\) \Comment{batch of input EEGs}
    \State \# I : \((batch, channel, height, width)\) \Comment{batch of input images}
    \Statex
    \State \# \(\tau\) : learned temperature parameter
    \State \# \(\beta\) : learned inner similarity parameter
    \State \# CS : Cosine Similarity
    \State \# SK : Similarity-Keeping
    \Statex
    \State \# extract normalized representations from the raw image and EEG
    \State $E_{f}$ = Norm(Linear($Enc_{eeg}$(E))) 
    \State $I_f$ = Norm($Enc_{img}$(I))  \Comment{can be obtained before training}
    \Statex
    \State \# calculate cosine similarity from the inner batch image and EEG
    \State $E_{CS}$ = CS($E_{f}$, $E_{f}$)
    \State $I_{CS}$ = CS($I_{f}$, $I_{f}$)
    \State $loss_{SK}$ = 1 - $\mathbb{E}$(CS($E_{CS}$, $I_{CS}$))
    \Statex
    \State \# scaled pairwise cosine similarity 
    \State logits = dot($E_{f}$, $I_f$.t) $\times$ $e^{\tau}$
    \Statex
    \State \# symmetric loss function
    \State labels = arange(batch)
    \Comment{self-supervised learning label}
    \State $loss_e$ = CrossEntropyLoss(logits, labels, axis=0)
    \State $loss_i$ = CrossEntropyLoss(logits, labels, axis=1)
    \State $total\_loss$ = ($loss_e$ + $loss_i$) / 2 + $\beta$ $\times$ $loss_{SK}$
\end{algorithmic}}
\end{algorithm}


\subsection{Network Architecture}

\subsubsection{EEG Encoder}

In this study, we introduce a series of multivariate time-series encoding architectures tailored to capture essential features in EEG data. Recent works suggest that upstream spatial convolution serves as an effective spatial filtering method for enhancing feature extraction and noise suppression \cite{wei2019spatial, pan2022matt}. Herein, we present the Spatial-Temporal Convolution (STConv) module, which employs spatial convolution to denoise data by referencing between brain electrodes, followed by temporal convolution. Additionally, we extend the capabilities of the STConv and Temporal-Spatial Convolution (TSConv) modules by integrating an attention mechanism, leading to the development of a novel transformer-like EEG encoder, which we refer to as NervFormer. In line with Graph Attention Networks (GATs) principles, we employ the Graph Attention (GA) module (see Appendix) to iteratively refine the state of each node, conceptualized as electrodes, by leveraging the states of all other nodes \cite{velivckovic2017graph, brody2021attentive}. The architectures of the baseline and proposed EEG encoders are illustrated and compared in Figure~\ref{fig:model_compare}.


\subsubsection{Image Encoder}

For our implementation, we integrate the off-the-shelf CLIP-ViT model \cite{radford2021learning}, which has demonstrated exceptional performance in aligning image and text representations. This model, pre-trained on extensive datasets, captures intricate details and high-level semantic information from images, making it an ideal candidate for our contrastive learning framework.


\subsubsection{Similarity-Keeping Contrastive Learning}

Inspired by recent neuroscience findings of the cortical network of visual object representation \cite{bao2020map, she2024temporal}, we take the interplay between object categories into account and propose a novel regularized contrastive learning framework. The procedure is outlined in Algorithm \ref{alg:1}.

The ordinary contrastive learning uses InfoNCE loss given by \cite{oord2018representation, he2020momentum, radford2021learning}:

\begin{equation}
\mathcal{L}_{InfoNCE} = -\mathbb{E}\left[\log \frac{\exp(S_{E, I}/\tau)}{\sum_{k=1}^N \exp(S_{E, I_k}/\tau)}\right]
\end{equation}

where the \(S_{E, I}\) denotes the similarity score between EEG signal \(E\) and image \(I\) pairing data, the \(\tau\) is learned temperature parameter, the training process shown in Figure~\ref{fig:music-plot}.

We introduce regularization to the ordinary contrastive learning by incorporating similarity preservation into the contrastive loss to capture both inter-sample and multimodal similarities. Drawing inspiration from the similarity-keeping (SK) concept used in knowledge distillation between EEG models \cite{huang2023enhancing}, we propose a novel SK loss to regularize the InfoNCE loss. This involves estimating the inner-batch inter-sample relationship.
The SK loss is defined as:

\begin{equation}
\mathcal{L}_{SK} = 1 -\mathbb{E}\left[S(S_{E, E}, S_{I, I})\right]
\end{equation}

We introduce a trainable parameter \(\beta\) to enhance training flexibility. When the \(\beta = 0\), the similarity-keeping InfoNCE loss reduces to the standard InfoNCE loss. The combined loss function, which integrates similarity-keeping, is illustrated in Figure~\ref{fig:sim_plot} and defined as:

\begin{equation}
\mathcal{L}_{SK-InfoNCE} = \mathcal{L}_{InfoNCE} + \beta \times \mathcal{L}_{SK}
\end{equation}

This integration of similarity-keeping into the contrastive loss framework ensures that the model not only aligns paired EEG and image embeddings effectively but also maintains the intrinsic relationships within the batch.

\section{Experiments}
\label{others}
\subsection{Datasets and Preprocessing}
The ThingsEEG dataset \cite{gifford2022large} comprises extensive EEG recordings gathered through a rapid serial visual presentation (RSVP) paradigm, featuring responses from 10 individuals to 16,740 natural images from the THINGS database \cite{hebart2019things}. The dataset includes 1654 training classes, each with 10 images, and 200 test classes, each with 1 image. EEG recordings were conducted using 64-channel EASYCAP equipment, and the data were preprocessed by segmenting into trials from 0 to 1000 ms post-stimulus onset, with baseline correction using the pre-stimulus mean. EEG responses for each image were averaged across repetitions, and the images were resized to 224×224 and normalized prior to processing.

\subsection{Experiment Details}
Experiments were conducted on a GeForce RTX 3090 24G GPU with Pytorch. Training using the MUSE series required approximately 2 to 3 hours per subject, with a batch size of 1000, while NervFormer series models took 40 minutes to 1 hour per subject. Models were saved at 200 epochs when the validation loss reached its lowest point. We use the weighted Adam optimizer with a learning rate of 0.0002 and parameters \(\beta_1\)=0.5 and \(\beta_2\)=0.999. The \(\tau\) in contrastive learning initialized with $log(1/0.07)$ and \(\beta\)=1. The NervFormer model achieves the best results with a multiheads number of 5. 
Results were averaged over five random seeds, and statistical significance was determined using the Wilcoxon Signed-Rank Test.

\subsection{Performance Comparison}

The comparison results presented in Table~\ref{tab:per_com} highlight the performance of various methods, with detailed model abbreviations provided in the appendix. 
Overall, MUSE-SK achieves the highest average top-1 accuracy at 19.3\%, while MUSE attains the highest average top-5 accuracy at 48.9\%. Furthermore, MUSE-SK-Nerv-GA, MUSE-Nerv-GA, MUSE, MUSE-SK, MUSE-SK-GA, MUSE-GA, and MUSE-SK-Nerv-GA significantly outperform the NICE-GA model in both top-1 ($p<0.01$) and top-5 ($p<0.01$) accuracy.  Although individual performance can differ, MUSE-based methods usually do better than others. The GA and SK variants are particularly strong in this evaluation.

\begin{table}[t!]
  \caption{Overall accuracy (\%) of 200-way zero-shot classification using CLIP-ViT as image encoder: top-1 and top-5. The parts in bold represent the best results, while the underlined parts are the second best.}
  \label{tab:per_com}
  \centering
  \Huge
\resizebox{\linewidth}{!}{
  \begin{tabular}{lcccccccccccccccccccccc}
    \toprule
    & \multicolumn{2}{c}{Subject 1} & \multicolumn{2}{c}{Subject 2} & \multicolumn{2}{c}{Subject 3} & \multicolumn{2}{c}{Subject 4} & \multicolumn{2}{c}{Subject 5} & \multicolumn{2}{c}{Subject 6} & \multicolumn{2}{c}{Subject 7} & \multicolumn{2}{c}{Subject 8} & \multicolumn{2}{c}{Subject 9} & \multicolumn{2}{c}{Subject 10} & \multicolumn{2}{c}{Ave} \\
    \cmidrule(r){2-3} \cmidrule(r){4-5} \cmidrule(r){6-7} \cmidrule(r){8-9} \cmidrule(r){10-11} \cmidrule(r){12-13} \cmidrule(r){14-15} \cmidrule(r){16-17} \cmidrule(r){18-19} \cmidrule(r){20-21} \cmidrule(r){22-23}
    Method & top-1 & top-5 & top-1 & top-5 & top-1 & top-5 & top-1 & top-5 & top-1 & top-5 & top-1 & top-5 & top-1 & top-5 & top-1 & top-5 & top-1 & top-5 & top-1 & top-5 & top-1 & top-5 \\
    \midrule
    \multicolumn{23}{c}{Subject dependent - train and test on one subject} \\
    \midrule
    BraVL & 6.1 & 17.9 & 4.9 & 14.9 & 5.6 & 17.4 & 5.0 & 15.1 & 4.0 & 13.4 & 6.0 & 18.2 & 6.5 & 20.4 & 8.8 & 23.7 & 4.3 & 14.0 & 7.0 & 19.7 & 5.8 & 17.5 \\   
    NICE & 12.3 & 36.6 & 10.4 & 33.9 & 13.1 & 39.0 & 16.4 & 47.0 & 8.0 & 26.9 & 14.1 & 40.6 & 15.2 & 42.1 & 20.0 & 49.9 & 13.3 & 37.1 & 14.9 & 41.9 & 13.8 & 39.5 \\
    NICE-SA & 13.3 & \uline{40.2} & 12.1 & 36.1 & 15.3 & 39.6 & 15.9 & 49.0 & 9.8 & 34.4 & 14.2 & 42.4 & 17.9 & 43.6 & 18.2 & 50.2 & 14.4 & 38.7 & 16.0 & 42.8 & 14.7 & 41.7 \\
    NICE-GA & \uline{15.2} & 40.1 & 13.9 & 40.1 & 14.7 & 42.7 & 17.6 & 48.9 & 9.0 & 29.7 & 16.4 & 44.4 & 14.9 & 43.1 & 20.3 & 52.1 & 14.1 & 39.7 & 19.6 & 46.7 & 15.6 & 42.8 \\
    \midrule
    MUSE-Nerv (ours)& 11.0 & 33.9 & 12.3 & 37.4 & 13.6 & 39.4 & 19.1 & 48.0 & 10.7 & 31.9 & 14.0 & 41.2 & 13.0 & 41.3 & 21.0 & 54.6 & 15.4 & 38.6 & 17.1 & 43.9 & 14.7 & 41.0\\
    MUSE-SK-Nerv (ours)& 11.6 & 34.7 & 14.3 & 40.4 & 13.6 & 38.2 & 20.8 & 48.6 & 12.0 & 32.2 & 16.1 & 41.5 & 15.7 & 43.7 & 24.1 & 54.4 & 17.2 & 41.7 & 17.1 & 44.7 & 16.3 & 42.0 \\
    MUSE-SK-Nerv-GA (ours)& 12.1 & 38.7 & 15.2 & 43.0 & 18.5 & 48.8 & 24.4 & 50.6 & \uline{14.0} & \textbf{36.6} & 18.0 & 46.1 & 19.7 & 48.4& 24.3 & 56.9 & \uline{17.8} & 43.7 & 21.9 & 52.2 & 18.6 & 46.5\\
    MUSE-Nerv-GA (ours) & 13.4 & 39.0 & \uline{17.6} & 42.8 & 17.3 & 48.0 & 22.6 & 50.3 & \textbf{14.4} & 35.9 & \uline{18.7} & 46.2 & 19.2 & 47.3 & \textbf{26.8} & 56.7 & \textbf{19.0} & \textbf{47.3} & 20.6 & 52.9 & 19.0 & 46.6 \\
    \midrule
    MUSE (ours)& 14.7 & 39.2 & 15.2 & \uline{45.3} & 19.3 & 48.7 & \uline{25.9} & \textbf{61.0} & 12.6 & \uline{36.0} & 18.5 & \uline{50.6} & \textbf{20.2} & \textbf{50.1} & \uline{26.3} & \uline{58.6} & \textbf{19.0} & 45.7 & 20.4 & 54.0 & \uline{19.2} & \textbf{48.9} \\
    MUSE-GA (ours) & 14.7 & 38.3 & 17.5 & \textbf{47.4} & 17.1 & 48.0 & 24.8 & 58.2 & 11.5 & 34.9 & 18.5 & 50.5 & 19.3 & 49.1 & 24.3 & 55.1 & 16.9 & 40.3 & \textbf{24.0} & \textbf{55.8} & 18.8 & 47.8\\
    MUSE-SK (ours)& 14.4 & 39.9 & 16.5 & 44.2 & \uline{19.7} & \uline{49.5} & \textbf{26.4} & \uline{58.6} & 13.2 & 34.0 & \textbf{19.1} & \textbf{52.5} & 19.5 & \uline{49.4} & \textbf{26.8} & \textbf{59.3} & 17.6 & \uline{46.6} & 20.1 & 54.3 & \textbf{19.3} & \uline{48.8}\\
    MUSE-SK-GA (ours)& \textbf{15.3} & \textbf{41.0} & \textbf{18.1} & 44.5 & \textbf{20.0} & \textbf{50.0} & 25.3 & 58.1 & 11.2 & 34.7 & 17.9 & 48.0 & \uline{20.1} & 49.1 & 25.4 & 57.7 & 17.0 & 43.6 & \uline{22.7} & \uline{54.4} & \textbf{19.3} & 48.1\\

    \bottomrule
  \end{tabular}}
\end{table}

\subsection{Ablation Study}

We conduct ablation studies on both MUSE and MUSE-Nerv series models, with the results of MUSE-Nerv illustrated in Table~\ref{tab:nervabla}. While the NervFormer EEG encoder does not demonstrate the best average zero-shot performance across all datasets, the MUSE-SK-Nerv-GA model achieves higher individual accuracy for subjects 5 and 10 compared to both MUSE and MUSE-SK. Moreover, beyond the MUSE series models, which solely employ the STConv as the EEG encoder, the MUSE-Nerv series models, incorporating the NervFormer as the EEG encoder, independently validate the efficacy of the similarity-keeping loss architecture and the graph attention module in EEG-image multimodal contrastive learning.

Upon examining the performance metrics of MUSE as depicted in Table~\ref{tab:3}, it becomes apparent that MUSE, MUSE-SK, and MUSE-SK-GA exhibit similar average performance levels. However, each method demonstrates distinct advantages across the ten subjects studied. For example, MUSE-SK-GA demonstrates superior overall performance in subjects 1, 3, and 10, while MUSE-SK achieves state-of-the-art results in subject 8. Additionally, each method excels uniquely in either top-1 or top-5 rankings in various subjects. This underscores the effectiveness of the SK and GA techniques as enhancements. However, in the context of STConv, these techniques do not demonstrate as clear an advantage as NervFormer does. We also observe that while SK may impact GA performance on NervFormer, both SK and GA enhance performance on STConv, with further details discussed in the model interpretation section.

\begin{table}
  \caption{Ablation Study of MUSE series models, accuracy (\%) of 200-way zero-shot classification: top-1 and top-5. The parts in bold represent the best results, while the underlined parts are the second best.}
  \label{tab:3}
  \centering
  \Huge
\resizebox{\linewidth}{!}{
  \begin{tabular}{lccccccccccccccccccccccc}
    \toprule
    & \multicolumn{2}{c}{Subject 1} & \multicolumn{2}{c}{Subject 2} & \multicolumn{2}{c}{Subject 3} & \multicolumn{2}{c}{Subject 4} & \multicolumn{2}{c}{Subject 5} & \multicolumn{2}{c}{Subject 6} & \multicolumn{2}{c}{Subject 7} & \multicolumn{2}{c}{Subject 8} & \multicolumn{2}{c}{Subject 9} & \multicolumn{2}{c}{Subject 10} & \multicolumn{2}{c}{Ave} & \multicolumn{1}{c}{Win}\\
    \cmidrule(r){2-3} \cmidrule(r){4-5} \cmidrule(r){6-7} \cmidrule(r){8-9} \cmidrule(r){10-11} \cmidrule(r){12-13} \cmidrule(r){14-15} \cmidrule(r){16-17} \cmidrule(r){18-19} \cmidrule(r){20-21} \cmidrule(r){22-23} \cmidrule(r){24-24}
    Method & top-1 & top-5 & top-1 & top-5 & top-1 & top-5 & top-1 & top-5 & top-1 & top-5 & top-1 & top-5 & top-1 & top-5 & top-1 & top-5 & top-1 & top-5 & top-1 & top-5 & top-1 & top-5 & subject score \#\\
    \midrule
    \multicolumn{23}{c}{Subject dependent - train and test on one subject} \\
    \midrule
    \multicolumn{23}{l}{\raggedright \textit{Original MUSE (STConv as EEG encoder \& CLIP-ViT as image encoder with InfoNCE loss)}} \\
     \textbf{MUSE} & \uline{14.7} & 39.2 & 15.2 & \textbf{45.3} & 19.3 & 48.7 & \uline{25.9} & \textbf{61.0} & \uline{12.6} & \textbf{36.0} & \uline{18.5} & \uline{50.6} & \textbf{20.2} & \textbf{50.1} & \uline{26.3} & \uline{58.6} & \textbf{19.0} & \uline{45.7} & \uline{20.4} & 54.0 & \uline{19.2} & \textbf{48.9} & \uline{6/20}\\
    \midrule
    \multicolumn{23}{l}{\raggedright \textit{Change InfoNCE loss to SK-InfoNCE loss}} \\
    \textbf{MUSE-SK} & 14.4 & \uline{39.9} & \uline{16.5} & 44.2 & \uline{19.7} & \uline{49.5} & \textbf{26.4} & \uline{58.6} & \textbf{13.2} & \uline{34.0} & \textbf{19.1} & \textbf{52.5} & 19.5 & \uline{49.4} & \textbf{26.8} & \textbf{59.3} & \uline{17.6} & \textbf{46.6} & 20.1 & \uline{54.3} & \textbf{19.3} & \uline{48.8} & \textbf{7/20}\\
    \midrule
    \multicolumn{23}{l}{\raggedright \textit{Change STConv to STConv-GA}} \\
    \textbf{MUSE-SK-GA} & \textbf{15.3} & \textbf{41.0} & \textbf{18.1} & \uline{44.5} & \textbf{20.0} & \textbf{50.0} & 25.3 & 58.1 & 11.2 & 34.7 & 17.9 & 48.0 & \uline{20.1} & 49.1 & 25.4 & 57.7 & 17.0 & 43.6 & \textbf{22.7} & \textbf{54.4} & \textbf{19.3} & 48.1 & \textbf{7/20}\\
    \bottomrule
  \end{tabular}}
\end{table}

\begin{table}
  \caption{Ablation Study of MUSE-Nerv series models,  accuracy (\%) of 200-way zero-shot classification: top-1 and top-5. The parts in bold represent the best results, while the underlined parts are the second best.}
  \label{tab:nervabla}
  \centering
  \Huge
\resizebox{\linewidth}{!}{
  \begin{tabular}{lccccccccccccccccccccccc}
    \toprule
    & \multicolumn{2}{c}{Subject 1} & \multicolumn{2}{c}{Subject 2} & \multicolumn{2}{c}{Subject 3} & \multicolumn{2}{c}{Subject 4} & \multicolumn{2}{c}{Subject 5} & \multicolumn{2}{c}{Subject 6} & \multicolumn{2}{c}{Subject 7} & \multicolumn{2}{c}{Subject 8} & \multicolumn{2}{c}{Subject 9} & \multicolumn{2}{c}{Subject 10} & \multicolumn{2}{c}{Ave} & \multicolumn{1}{c}{Win}\\
    \cmidrule(r){2-3} \cmidrule(r){4-5} \cmidrule(r){6-7} \cmidrule(r){8-9} \cmidrule(r){10-11} \cmidrule(r){12-13} \cmidrule(r){14-15} \cmidrule(r){16-17} \cmidrule(r){18-19} \cmidrule(r){20-21} \cmidrule(r){22-23} \cmidrule(r){24-24}
    Method & top-1 & top-5 & top-1 & top-5 & top-1 & top-5 & top-1 & top-5 & top-1 & top-5 & top-1 & top-5 & top-1 & top-5 & top-1 & top-5 & top-1 & top-5 & top-1 & top-5 & top-1 & top-5 & subject score \#\\
    \midrule
    \multicolumn{24}{c}{Subject dependent - train and test on one subject} \\
    \midrule
    \multicolumn{24}{l}{\raggedright \textit{Original MUSE-Nerv (NervFormer as EEG encoder \& CLIP-ViT as image encoder with InfoNCE loss)}} \\
    \textbf{MUSE-Nerv} & 11.0 & 33.9 & 12.3 & 37.4 & \uline{13.6} & \uline{39.4} & 19.1 & 48.0 & 10.7 & 31.9 & 14.0 & 41.2 & 13.0 & 41.3 & 21.0 & 54.6 & 15.4 & 38.6 & \uline{17.1} & 43.9 & 14.7 & 41.0 & 0\\
    \midrule
    \multicolumn{24}{l}{\raggedright \textit{Change InfoNCE loss to SK-InfoNCE loss}} \\
    \textbf{MUSE-SK-Nerv} & \uline{11.6} & \uline{34.7} & \uline{14.3} & \uline{40.4} & \uline{13.6} & 38.2 & \uline{20.8} & \uline{48.6} & \uline{12.0} & \uline{32.2} & \uline{16.1} & \uline{41.5} & \uline{15.7} & \uline{43.7} & \uline{24.1} & \uline{54.4} & \uline{17.2} & \uline{41.7} & \uline{17.1} & \uline{44.7} & \uline{16.3} & \uline{42.0} & 0\\
    \midrule
    \multicolumn{24}{l}{\raggedright \textit{Change NervFormer to NervFormer-GA}} \\
    \textbf{MUSE-SK-Nerv-GA} & \textbf{12.1} & \textbf{38.7} & \textbf{15.2} & \textbf{43.0} & \textbf{18.5} & \textbf{48.8} & \textbf{24.4} & \textbf{50.6} & \textbf{14.0} & \textbf{36.6} & \textbf{18.0} & \textbf{46.1} & \textbf{19.7} & \textbf{48.4} & \textbf{24.3} & \textbf{56.9} & \textbf{17.8} & \textbf{43.7} & \textbf{21.9} & \textbf{52.2} & \textbf{18.6} & \textbf{46.5} & \textbf{20/20}\\
    \bottomrule
  \end{tabular}}
\end{table}

\begin{figure}[ht]
    \centering
    \begin{minipage}{0.45\linewidth}
        \centering
        \includegraphics[width=\linewidth]{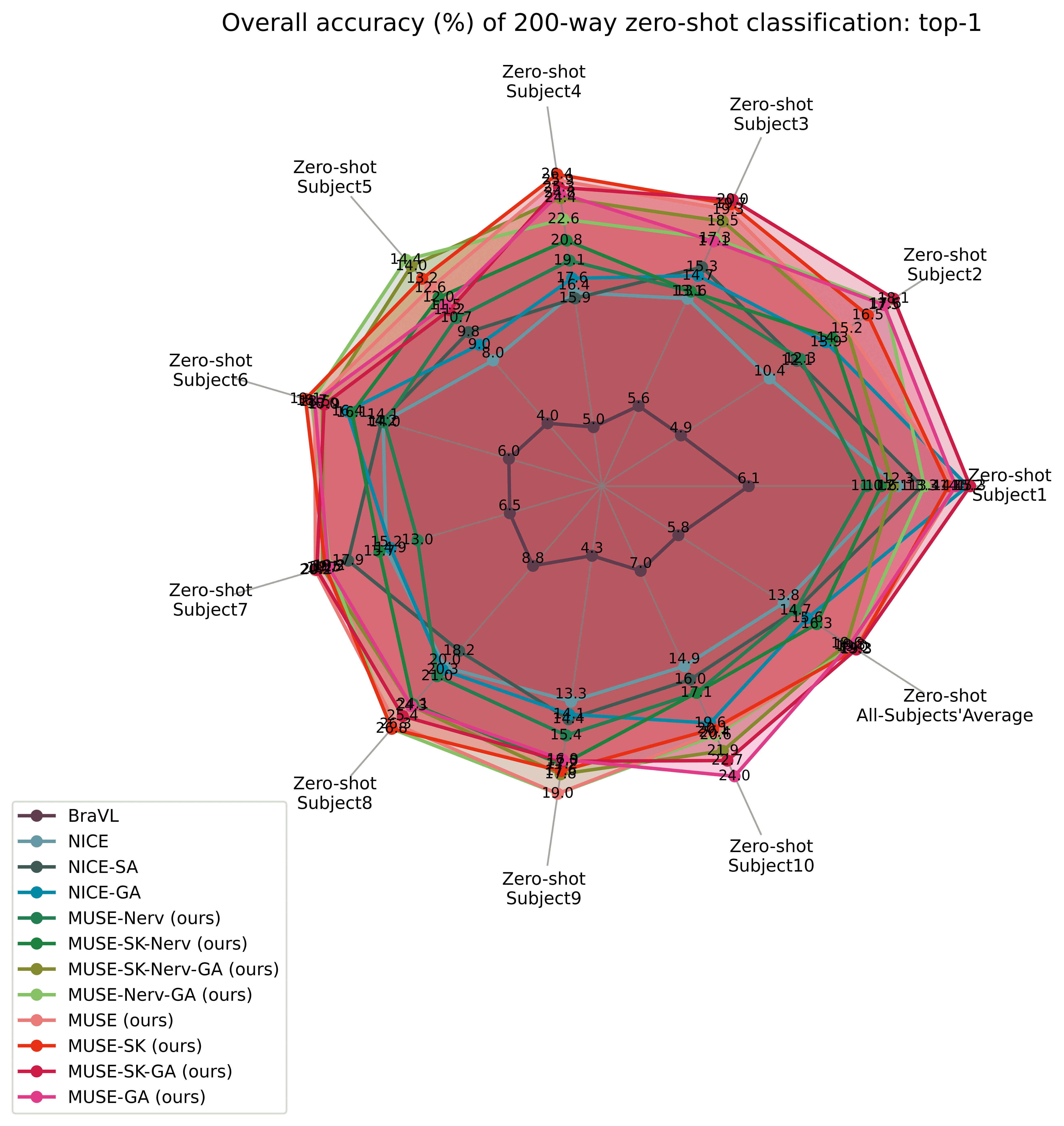}
        \caption{Overall Top-1 zero-shot accuracy comparison of all models.}
        \label{fig:top-1-acc}
    \end{minipage}
    \hfill
    \begin{minipage}{0.45\linewidth}
        \centering
        \includegraphics[width=\linewidth]{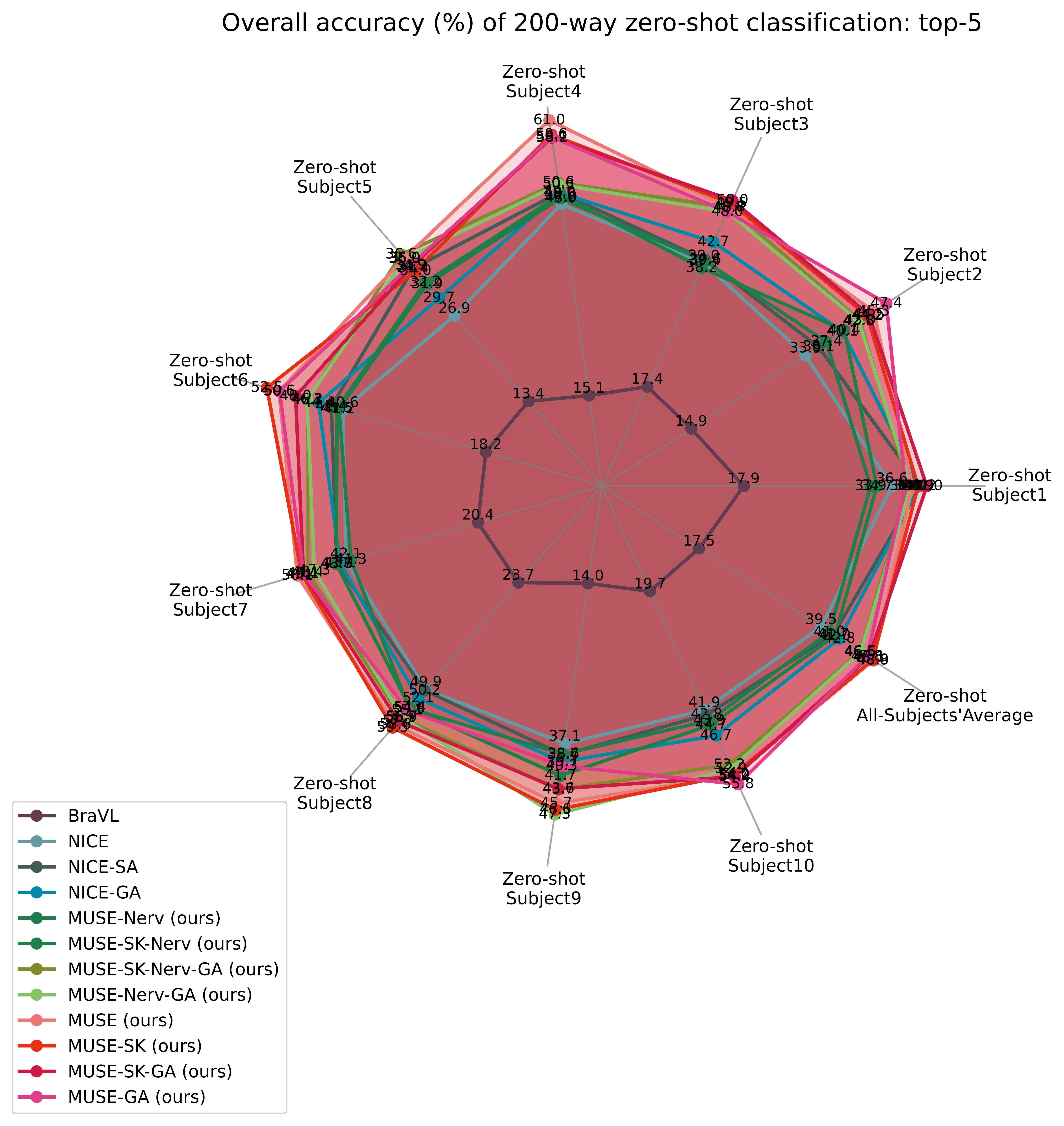}
        \caption{Overall Top-5 zero-shot accuracy comparison of all models.}
        \label{fig:top-5-acc}
    \end{minipage}
\end{figure}

\subsection{Model Interpretation}
We conducted model interpretation to uncover the internal mechanisms of our models across three distinct domains: spatial-temporal, brain region topography-temporal, and temporal-frequency. We employed the Grad-CAM analysis method \cite{selvaraju2016grad} to scrutinize our proposed best MUSE series models.

\subsubsection{Spatial-Temporal Dynamics Analysis}

\begin{figure}
    \centering
    \includegraphics[width=.9\linewidth]{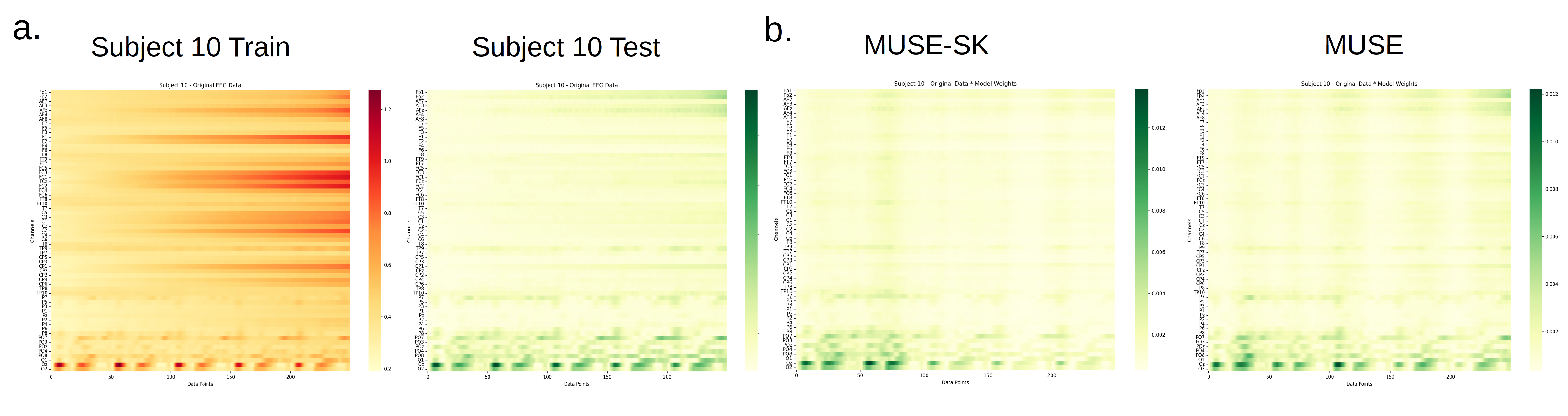}
    \caption{(a) Grad-CAM visualization of the MUSE series model averaged across all trials and repetitions for subject 10. (b) Comparative analysis reveals that MUSE-SK exhibits a heightened focus on the occipital lobes during the 100-500 ms time window compared to MUSE-SK and other models. }
    \label{fig:sub10_eeg_heat}
\end{figure}

\begin{figure}
    \centering
    \includegraphics[width=1\linewidth]{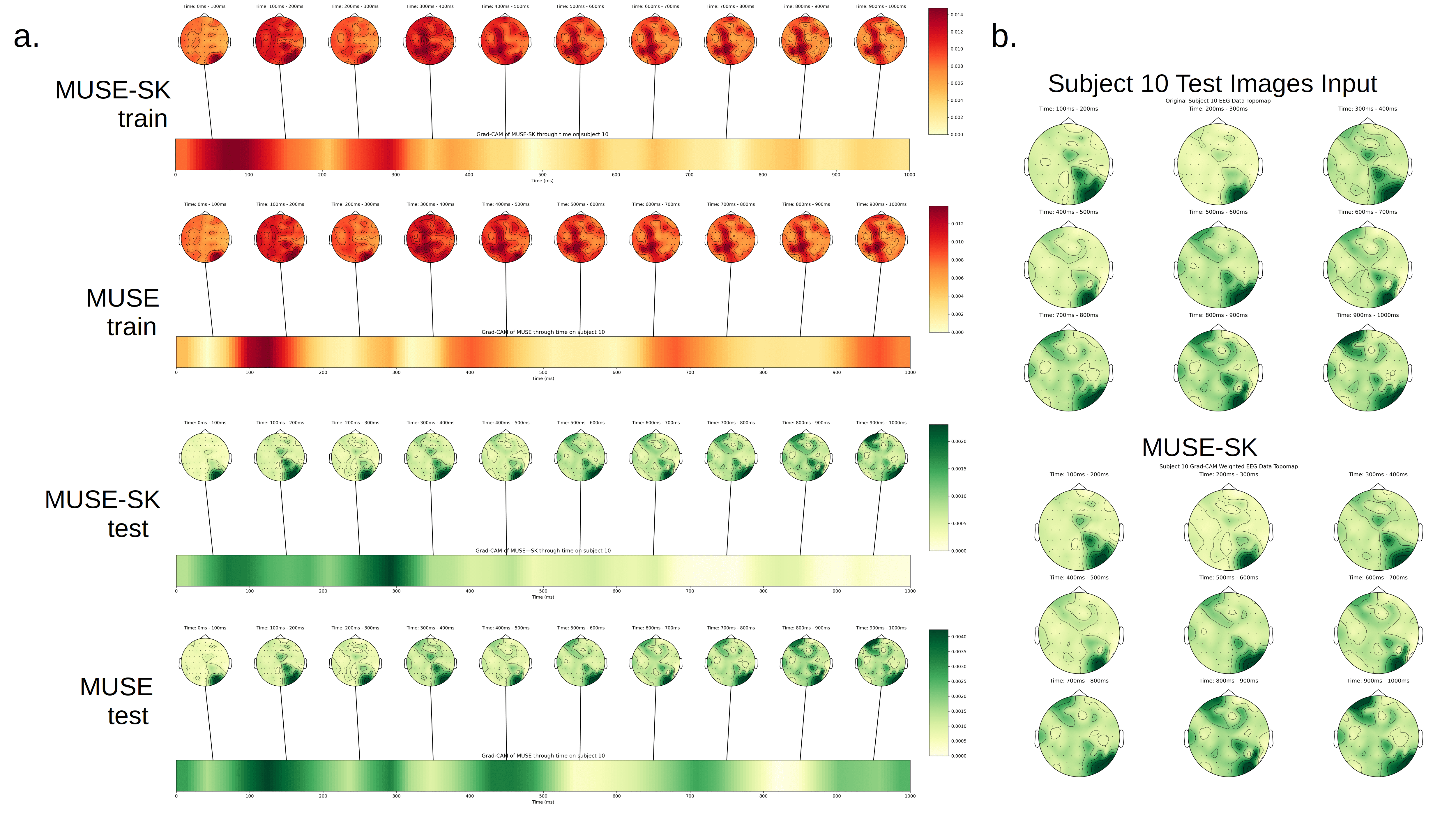}
    \caption{Topomap depicting the average response over each 100 ms interval across all trials, aggregated over all repetitions for subject 10. (a) Grad-CAM visualization for both MUSE-SK and MUSE models is presented, with the color bar at the bottom indicating the intensity of Grad-CAM for each model over time. Both models predominantly focus on the 100-500 ms time window. (b) A zoomed-in comparison between the input EEG data and the MUSE-SK model highlights the model's enhanced focus on temporal and occipital areas.}
    \label{fig:sub10_MUSE_com}
\end{figure}

To ensure that meaningful signals are preserved during Grad-CAM calculations, we take the absolute value of all Grad-CAM and EEG signal intensities of each trial for further analysis.
The spatial-temporal comparison on both training and testing trials is depicted in Figure~\ref{fig:sub10_eeg_heat}. We note that the higher-performing models, such as MUSE-SK and MUSE-SK-GA, concentrate on the EEG information between the 25th and 125th data points, corresponding to the 100 ms to 500 ms time period. Figure~\ref{fig:sub10_MUSE_com} illustrates a distinct response observed in the occipital cortex between 100 and 600 ms after the onset in MUSE-SK. However, the 200 ms stimulus onset asynchrony (SOA) continues to elicit periodic responses in the occipital cortex. Furthermore, a response in the parietal cortex is evident after 100 ms. This observation aligns with the bottom-up hierarchy of the visual system \cite{dicarlo2007untangling}, wherein visual stimuli are sequentially processed by V1, V2, and V4 in the occipital cortex, and subsequently by the inferotemporal region in the temporal cortex along the ventral stream for object recognition \cite{bao2020map}.


\section{Conclusion}

In summary, this paper introduces the MUltimodal Similarity-keeping contrastivE learning (MUSE) framework, a novel approach tailored specifically for zero-shot EEG-based image classification, thereby addressing the intricate challenge of deciphering visual information from non-invasive EEG signals.
Our method, drawing inspiration from established neuroscience findings, achieves state-of-the-art decoding accuracy, as substantiated by rigorous experimental evaluations. We further interpret our models and uncover insights into the spatial-temporal dynamics of EEG responses, shedding light on the neural processes underlying visual perception. We foresee that our work will catalyze further exploration in bridging the gap between EEG decoding and image recognition, advancing our understanding of visual cognition in the human brain.

{
\small
\newpage
\bibliographystyle{unsrt}
\bibliography{neurips_2024}
\newpage
\appendix

\section{Appendix}
\subsection{The model abbreviations details}
The abbreviations detail is shown as Table~\ref{tab:model_label}. 
\label{sec:appendA0}
\begin{table}[ht]
    \caption{The detail of all the model}
    \centering
    \begin{tabular}{cccc}
    \toprule
      Method  & EEG Encoder & Image Encoder & Loss Function\\
    \midrule
    \midrule
      BraVL \cite{du2023decoding} & MLP & MLP & ELBO \\
    \midrule
    \midrule
      NICE \cite{song2023decoding}   & TSConv  & CLIP-ViT & InfoNCE \\
      NICE-SA \cite{song2023decoding} & TSConv-SA & CLIP-ViT & InfoNCE \\
      NICE-GA \cite{song2023decoding} & TSConv-GA & CLIP-ViT  & InfoNCE \\
      \midrule
      MUSE (ours)  & STConv & CLIP-ViT & InfoNCE \\
      MUSE-GA (ours)  & STConv-GA & CLIP-ViT & InfoNCE \\
      MUSE-Nerv (ours)  & NervFormer & CLIP-ViT & InfoNCE \\
      MUSE-Nerv-GA (ours)  & NervFormer-GA & CLIP-ViT & InfoNCE \\
      \midrule
      MUSE-SK (ours)  & STConv & CLIP-ViT & SK-InfoNCE \\
      MUSE-SK-GA (ours) & STConv-GA & CLIP-ViT & SK-InfoNCE \\
      MUSE-SK-Nerv (ours)  & NervFormer & CLIP-ViT & SK-InfoNCE \\
      MUSE-SK-Nerv-GA (ours)  & NervFormer-GA & CLIP-ViT & SK-InfoNCE \\
      
    \bottomrule
    \end{tabular}
    \label{tab:model_label}
\end{table}
\subsection{Graph Attention}
In line with Graph Attention Networks (GATs) principles, we employ the Graph Attention (GA) module to iteratively refine the state of each node, conceptualized as electrodes, by leveraging the states of all other nodes \cite{velivckovic2017graph, brody2021attentive}. Through these mechanisms, the GA module dynamically adjusts the importance of each node based on the contextual information proffered by its neighbors, ensuring an attention-weighted update that underscores the interconnectivity of node features within the graph's architecture. Each node's representation is denoted by \( n_i \in \mathbb{R}^{1 \times T} \), indexed by \( i \) for \( i = 1, \ldots, ch \), signifying an electrode that establishes connections with a defined set \( \mathcal{N}_i \) of adjacent nodes, thus forming a fully connected graph.
The update mechanism for an individual node \( n_i \) is formalized as:
\begin{equation}
n_i' = \alpha_{i,i}Wn_i + \sum_{j \in \mathcal{N}_i} \alpha_{i,j}Wn_j
\end{equation}
where \( n_i' \) designates the updated node, \( \alpha_{i,j} \) encapsulates the attention coefficients indicative of the feature significance from node \( j \) to node \( i \), and \( W \) is the weight matrix of the linear transformation.
The attention coefficients \( \alpha_{i,j} \) are computed via the equation:
\begin{equation}
\alpha_{i,j} = \frac{\exp(a^T \cdot \text{LeakyReLU}(W[n_i \| n_j]))}{\sum_{k \in \mathcal{N}_i \cup \{i\}} \exp(a^T \cdot \text{LeakyReLU}(W[n_i \| n_k]))}
\end{equation}
In this expression, \( a \in \mathbb{R}^{2T} \) represents the weight vector of a feedforward attention mechanism, \( ()^T \) indicates the transpose operation, and \( \| \) signifies concatenation. LeakyReLU is introduced as the non-linear function with a negative slope coefficient of 0.2, facilitating computational stability and non-linearity. 

\subsection{Time-Frequency Dynamics Analysis}
We took the best SK model, MUSE-SK, to perform time-frequency analysis and found that the alpha wave, gamma wave, and theta wave signals were concentrated on the occipital and parietal lobes in both the training and testing topomaps. This finding aligns with medical literature, where the alpha wave is associated with visual attention \cite{klimesch1999eeg, mathewson2011pulsed}, and the gamma wave is related to higher cognitive functions, attention, and visual processing \cite{fries2001modulation}. This also indicates that our designed model has indeed learned some neural behaviors related to the human brain.

\begin{figure}
    \centering
    \includegraphics[width=1\linewidth]{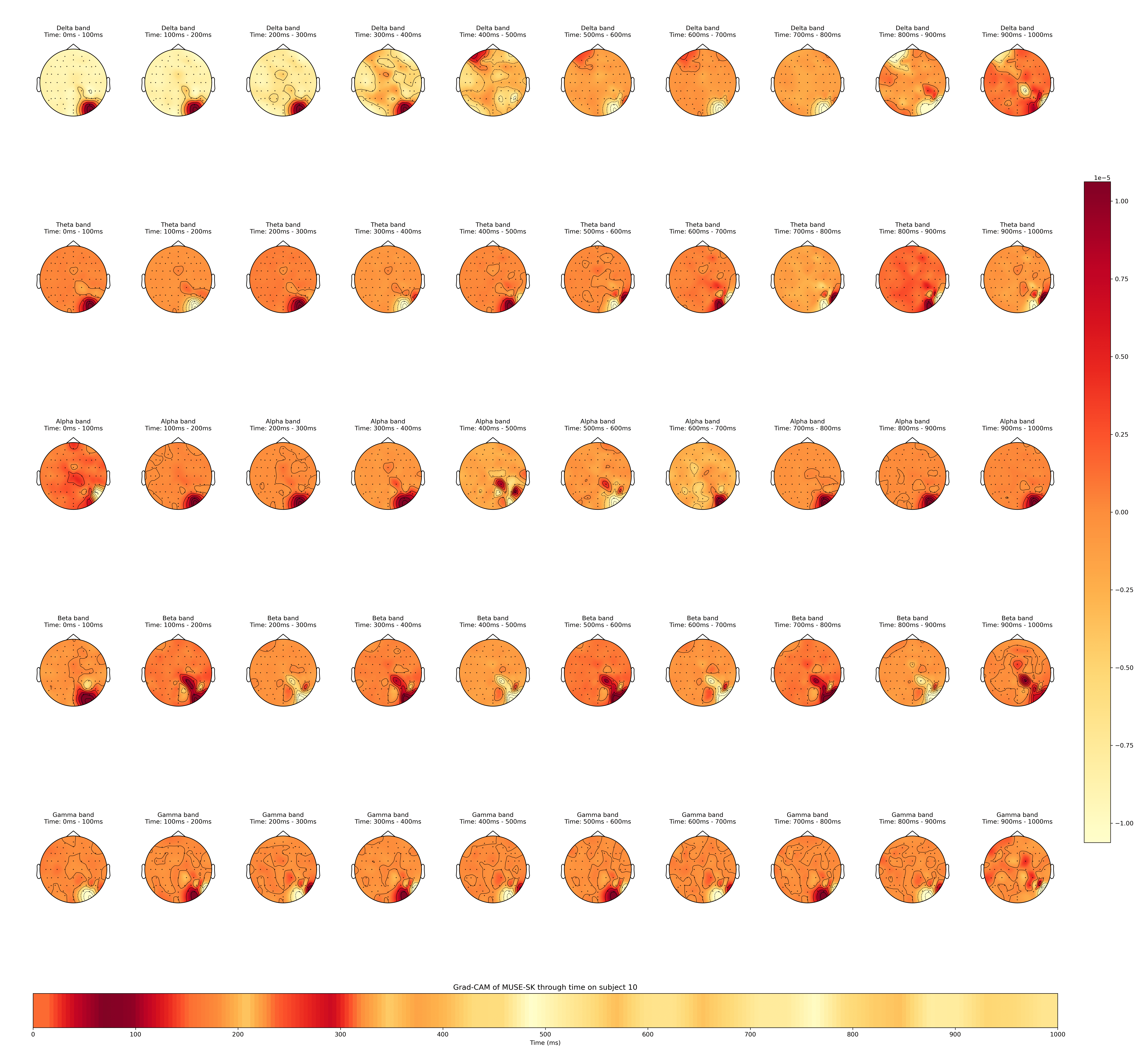}
    \caption{Time-Frequency map of MUSE-SK on averaging all of subject 10's training trials. We can see that the MUSE-SK can focus on alpha band and gamma band, where is related to vision attention and high-level visual recognition in neural science.}
    \label{fig:sub10_MUSE_SK_GA_time_feq_train}
\end{figure}

\begin{figure}
    \centering
    \includegraphics[width=1\linewidth]{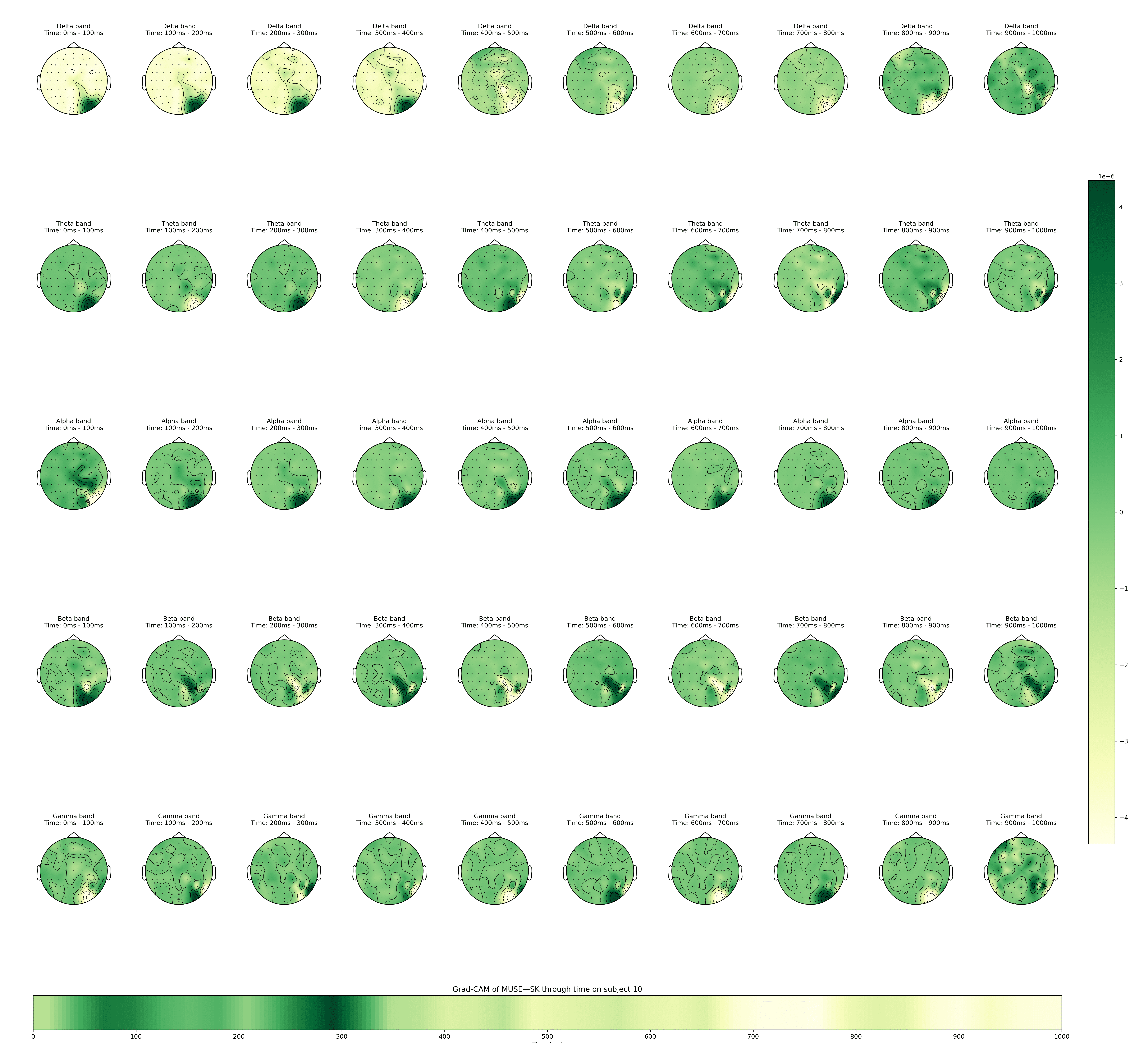}
    \caption{Time-Frequency map of MUSE-SK on averaging all trials in the testing set of subject 10. It is evident that MUSE-SK focuses on the alpha and gamma bands, which are associated with visual attention and high-level visual recognition in neuroscience.}
    \label{fig:sub10_MUSE_SK_GA_time_feq_test}
\end{figure}
\subsection{Limitation}
\label{Limitation}
In our framework, we have not changed the image encoder to the more powerful CLIP, but we focus on comparing different EEG encoders under the same image encoder and the reliability of our proposed brain-inspired similarity-keeping framework. After demonstrating that this work can indeed improve the performance of contrastive learning, replacing the image encoder with a more powerful one would be a better direction.

\subsection{Table of Testing Object Categories}
\label{Objectcat}

We also try to use Grad-CAM method doing model interpretation on testing sets with our-selected category.

 \begin{figure}[ht]
     \centering
     \includegraphics[width=1\linewidth]{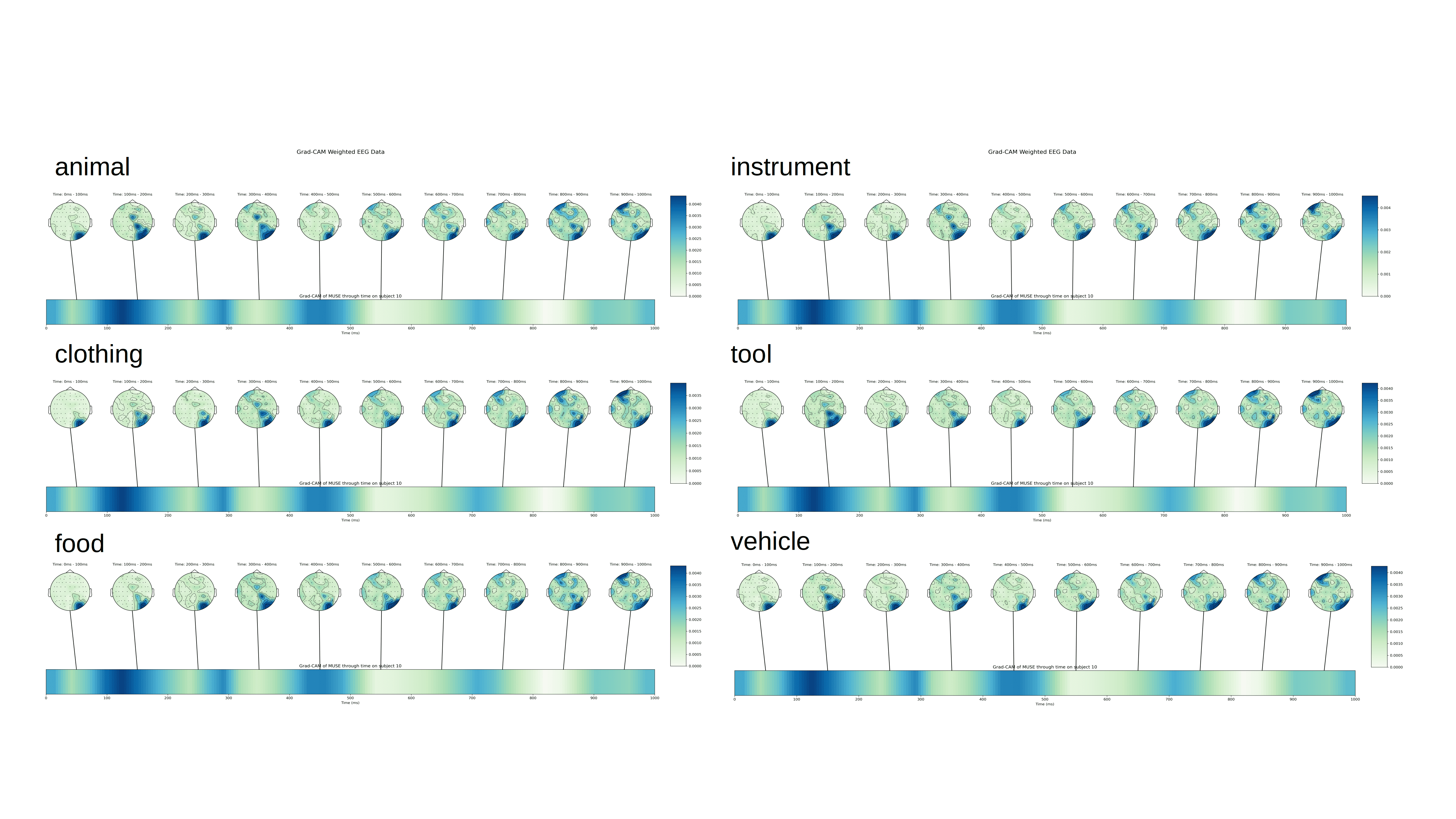}
     \caption{MUSE model interpretation on our-selected category.}
     \label{fig:enter-label}
 \end{figure}

\begin{table}[ht]
    \caption{Test images on THINGSEEG dataset categories}
    \centering
    \begin{tabular}{p{3cm} p{12cm}}
     \toprule
      \textbf{Category} & \textbf{Items} \\
    \midrule
     animal & 00002\_antelope, 00012\_beaver, 00024\_bug, 00033\_cat, 00034\_caterpillar, 
     00039\_cheetah, 00046\_cobra, 00053\_crab, 00058\_crow, 00063\_dalmatian, 00065\_dragonfly, 00069\_eagle, 00070\_eel, 00072\_elephant, 00076\_flamingo, 00086\_goose, 00087\_gopher, 00088\_gorilla, 00089\_grasshopper, 00097\_hummingbird, 00106\_lamb, 00110\_lightning\_bug, 00111\_manatee, 00117\_mosquito, 00127\_ostrich, 00129\_panther, 00133\_pheasant, 00136\_pigeon, 00137\_piglet, 00142\_possum, 00144\_pug, 00150\_rhinoceros, 00152\_rooster, 00161\_seagull, 00183\_tick, 00190\_turkey \\
     \midrule
     clothing & 00019\_bonnet, 00037\_chaps, 00043\_cleat, 00045\_coat, 00052\_coverall, 00074\_face\_mask, 00083\_glove, 00094\_headscarf, 00096\_hoodie, 00104\_kneepad, 00107\_lampshade, 00128\_pajamas, 00138\_pocket, 00155\_sandal, 00169\_snowshoe, 00176\_suit, 00177\_t-shirt, 00182\_tiara, 00187\_top\_hat, 00189\_tube\_top \\
     \midrule
     instruments & 00009\_bassoon, 00041\_chime, 00067\_drum, 00080\_french\_horn, 00119\_music\_box, 00149\_recorder\\
     \midrule
     food & 00005\_banana, 00007\_basil, 00011\_batter, 00015\_birthday\_cake, 00018\_bok\_choy, 00022\_bread, 00027\_bun, 00029\_calamari, 00032\_cashew, 00038\_cheese, 00047\_coconut, 00048\_coffee\_bean, 00050\_cookie, 00051\_cordon\_bleu, 00054\_creme\_brulee, 00055\_crepe, 00057\_croissant, 00060\_crumb, 00061\_cupcake, 00064\_dessert, 00071\_egg, 00073\_espresso, 00081\_fruit, 00082\_garlic, 00091\_hamburger, 00098\_ice\_cube, 00101\_jelly\_bean, 00109\_lettuce, 00112\_marijuana, 00113\_meatloaf, 00120\_mussel, 00122\_okra, 00123\_omelet, 00124\_onion, 00125\_orange, 00126\_orchid, 00131\_pear, 00132\_pepper1, 00135\_pie, 00140\_popcorn, 00141\_popsicle, 00143\_pretzel, 00147\_radish, 00148\_raspberry, 00157\_sausage, 00158\_scallion, 00159\_scallop, 00162\_seaweed, 00163\_seed, 00174\_strawberry, 00184\_tomato\_sauce, 00195\_walnut, 00196\_wheat, 00199\_wine \\
     \midrule
     tool & 00003\_backscratcher, 00006\_baseball\_bat, 00016\_blowtorch, 00020\_bottle\_opener, 00021\_brace, 00023\_breadbox, 00026\_bullet, 00030\_candlestick, 00035\_cd\_player, 00042\_chopsticks, 00044\_cleaver, 00049\_coffeemaker, 00062\_dagger, 00078\_fork, 00079\_freezer, 00090\_grenade, 00092\_hammer, 00093\_handbrake, 00103\_kettle, 00105\_ladle, 00114\_metal\_detector, 00118\_muff, 00130\_paperweight, 00134\_pickax, 00139\_pocketknife, 00145\_punch2, 00168\_slingshot, 00170\_spatula, 00171\_spoon, 00173\_stethoscope, 00185\_tongs, 00186\_tool, 00192\_vise, 00197\_wheelchair, 00200\_wok \\
    \midrule
     vehicle & 00001\_aircraft\_carrier, 00014\_bike, 00017\_boat, 00025\_buggy, 00031\_cart, 00059\_cruise\_ship, 00075\_ferry, 00084\_golf\_cart, 00085\_gondola, 00100\_jeep, 00115\_minivan, 00154\_sailboat, 00160\_scooter, 00164\_skateboard, 00165\_sled, 00172\_station\_wagon, 00175\_submarine, 00191\_unicycle \\
     \midrule
     other & Other categories in test images. \\
    \bottomrule
    \end{tabular}
    \label{tab:my_label}
\end{table}

\begin{figure}
    \centering
    \includegraphics[width=1\linewidth]{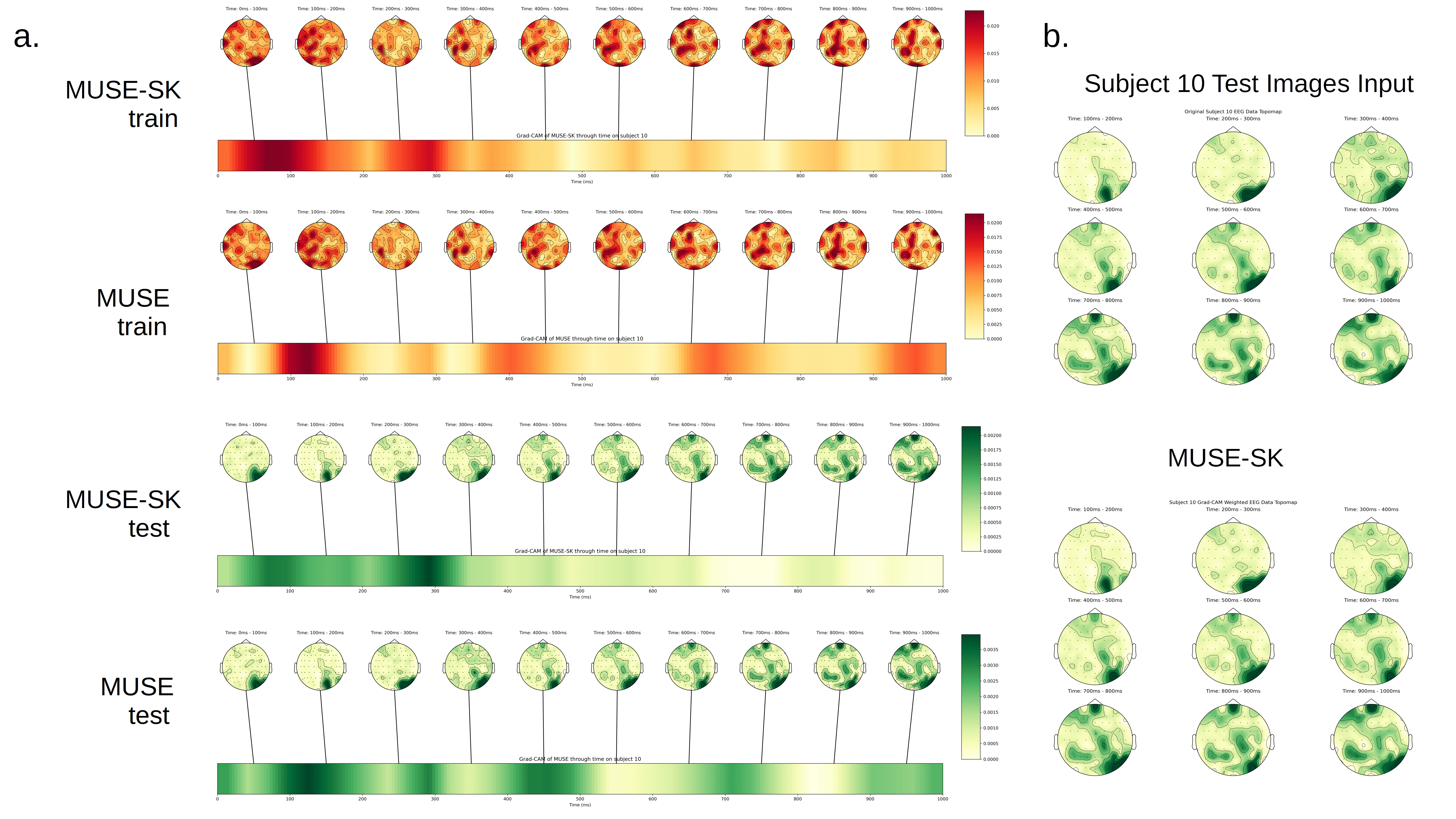}
    \caption{Topomap of each 100 ms by on one trial averaging through all the repetition on subject 10. (a.) On MUSE-SK and MUSE models, the color bar on the botton is the Grad-CAM of each model through time. Most of the model focus on the 100-500ms. The u (b.) Zoom-in and compare the input EEG data and the MUSE-SK, can see that the model can more focus on temporal and occipital areas.  }
    \label{fig:sub10_MUSE_com_one}
\end{figure}

\begin{figure}
    \centering
    \includegraphics[width=1\linewidth]{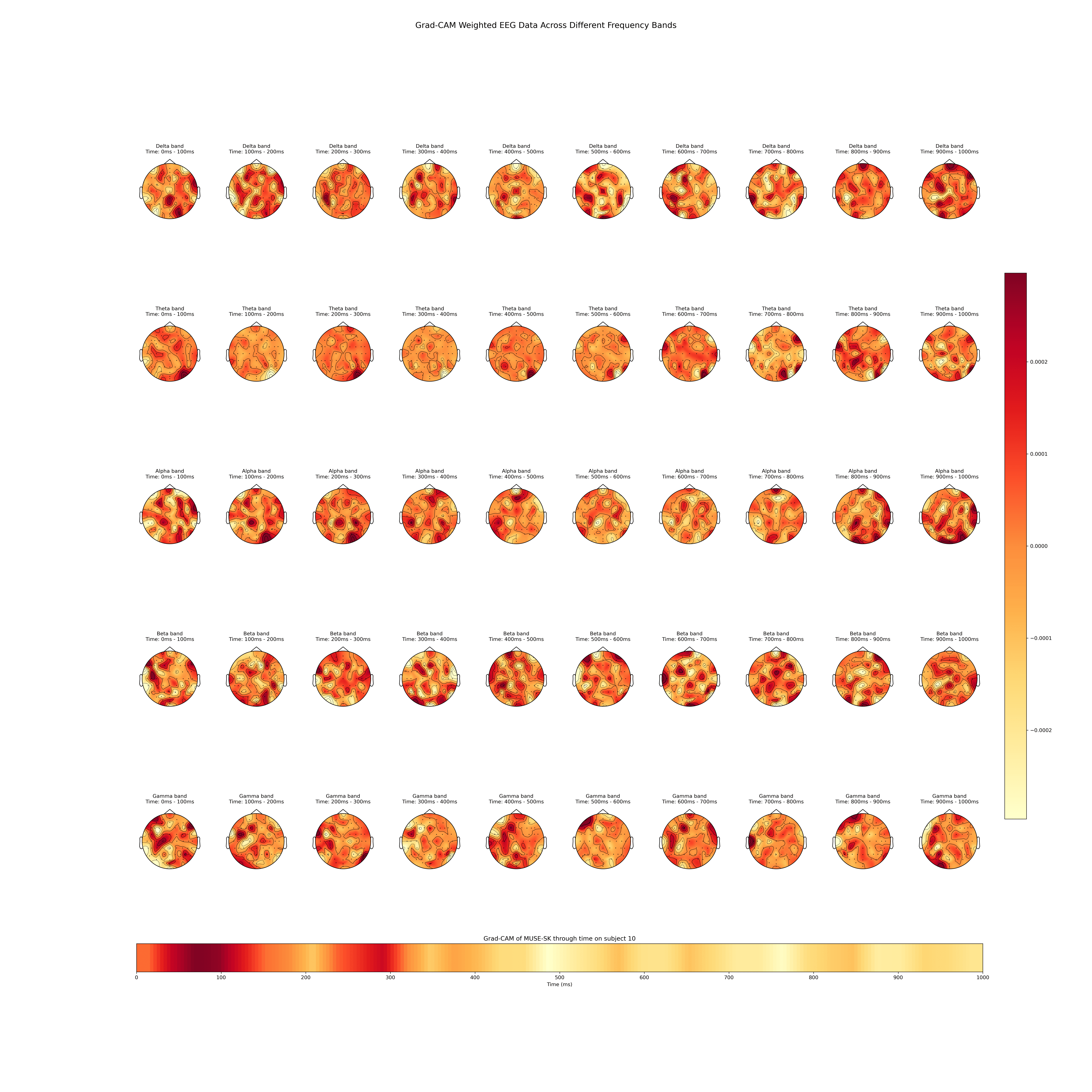}
    \caption{Time-Frequency map of MUSE-SK on one of subject 10's training trial. We can see that the MUSE-SK can focus on alpha band and gamma band, where is related to vision attention and high-level visual recognition in neural science.}
    \label{fig:sub10_MUSE_SK_GA_time_feq_train_one}
\end{figure}

\begin{figure}
    \centering
    \includegraphics[width=1\linewidth]{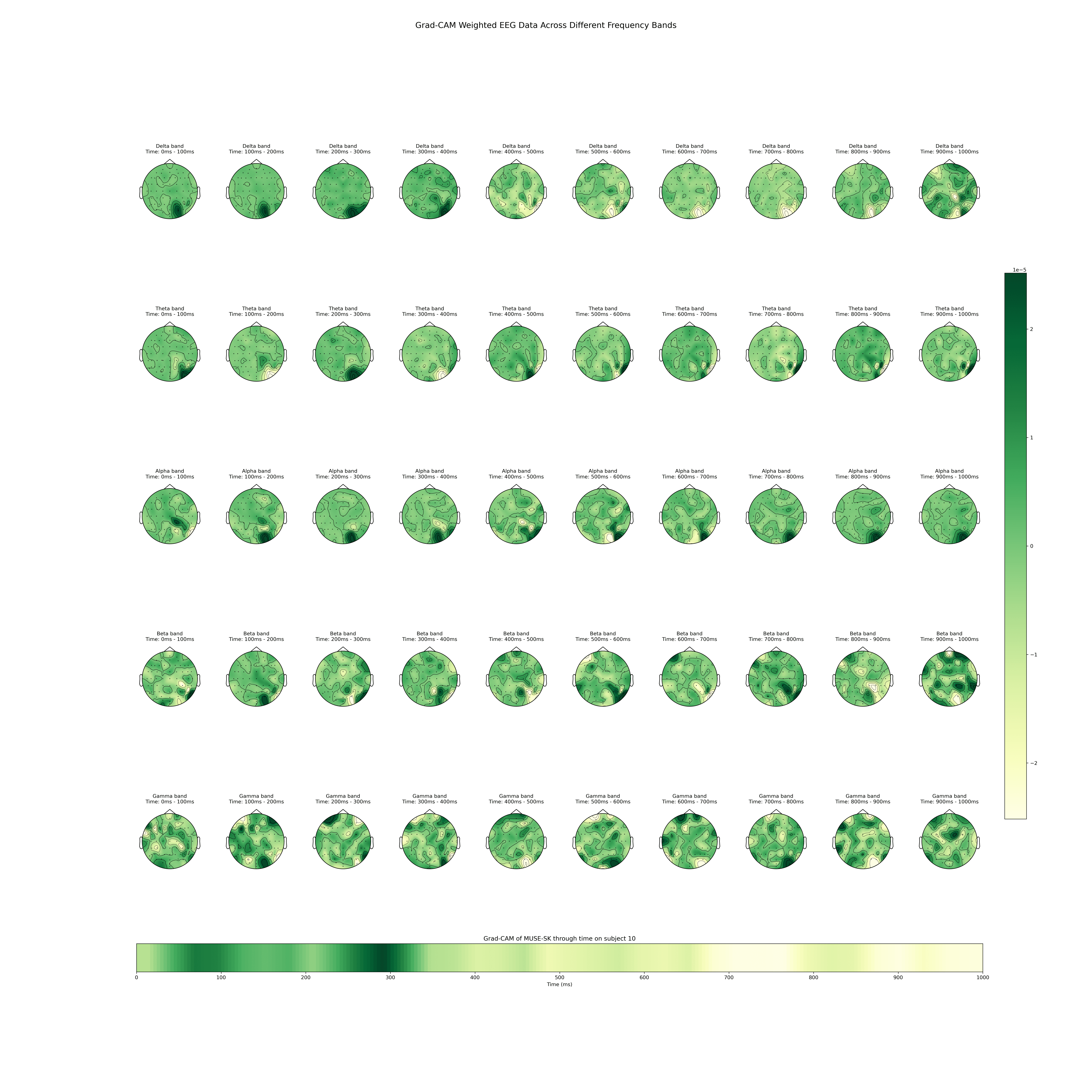}
    \caption{Time-Frequency map of MUSE-SK on one trial in the testing set of subject 10. It is evident that MUSE-SK focuses on the alpha and gamma bands, which are associated with visual attention and high-level visual recognition in neuroscience.}
    \label{fig:sub10_MUSE_SK_GA_time_feq_test_one}
\end{figure}

}

\end{document}